\documentclass[aps,prl,twocolumn,superscriptaddress]{revtex4-1}
\usepackage{bm}
\usepackage{amsfonts}
\usepackage{amssymb}
\usepackage{epsfig}
\usepackage{amsmath}
\usepackage{times}
\usepackage{color}
\usepackage{mathtools}
\usepackage[T1]{fontenc}
\usepackage{multirow,booktabs}
\usepackage{array}
\usepackage{soul}
\begin{document}
\title{Symmetry Hierarchy and Thermalization Frustration in Graphene Nanoresonators}
\author{Yisen Wang}
\affiliation{Lanzhou Center for Theoretical Physics, Key Laboratory of Theoretical Physics of Gansu Province, and Key Laboratory for Magnetism and Magnetic Materials of MOE, Lanzhou University, Lanzhou, Gansu 730000, China}

\author{Liang Huang}\thanks{Corresponding author: huangl@lzu.edu.cn}
\affiliation{Lanzhou Center for Theoretical Physics, Key Laboratory of Theoretical Physics of Gansu Province, and Key Laboratory for Magnetism and Magnetic Materials of MOE, Lanzhou University, Lanzhou, Gansu 730000, China}
\date{\today}

\begin{abstract}
As the essential cause of the intrinsic dissipation that limits the quality of graphene nanoresonators, intermodal energy transfer is also a key issue in thermalization dynamics. {Typically} systems with larger initial energy demand shorter time to be thermalized. {\color{black} However, we find quantitatively that} instead of becoming shorter, the equipartition time of the graphene nanoresonator can increase abruptly by one order of magnitude. This thermalization frustration emerges due to the partition of the normal modes based on the hierarchical symmetry, and a sensitive on-off switching of the energy flow channels between symmetry classes controlled by Mathieu instabilities. The results uncover the decisive roles of symmetry in the thermalization at the nanoscale, and may also lead to strategies for improving the performance of graphene nanoresonators.
\end{abstract}

\maketitle

{\it Introduction.}---
Graphene nanoresonators have attracted much attention in recent years because of their superb mechanical responses due to the extremely high quality factor for sensing applications \cite{Bunch2007,lee2008,NC_tian2018,yg_sensor,2014NatNa,2017NatCo, science_Klimov2012,Guttinger2017NN}.
The inherent intermodal energy transfer opens up additional energy dissipation channel other than the interaction with the environment and sets up an upper limit to the quality factor of graphene nanoresonators \cite{FPUprl}. Indeed, evidence of intermodal energy transfer has been reported from energy decay measurements in multilayer graphene resonators \cite{Guttinger2017NN}. Therefore engineering the intermodal interaction becomes critical for controlling internal energy dissipation. This, on the other hand, is exactly the core issue of thermalization. Thus graphene nanoresonators supply outstanding platforms for the investigation of thermalization at the nanoscale \cite{FPUprl,wang2018symmetry,wang2018pre}, which in turn can be exploited to improve their performance.

Thermalization and energy equipartition hypothesis are central issues of statistical physics \cite{WT_pnas,Lov2018PRL,Zhao2020PRL,Dhar2007PRL,Metastability_scenario, FPU_rev,twodimensional_FPU1,twodimensional_FPU2,Zhao2D2020, Mussot2014PRX,Conti_2013,Cassidy2009,Dahai2021}, which have been recently investigated in photonic lattices \cite{Christodoulides2003, LahiniAP2008,Parity2019,kon2015phot}, trapped-ion arrays \cite{RMP2019many,Detail2018,Time2016clos,LemmerCSS2015,DingMHM2017}, optical fibers \cite{muss2018fibre,Obser2018,Exp2007}, etc.
In nonlinear lattices, thermalization typically indicates energy equipartition among all the modes \cite{Benettin2008}.
The intermodal couplings, especially those with large values, build up the energy flow pathway and guide the system to thermalized state.
For example, when nonlinearity is weak, in the Fermi-Pasta-Ulam-Tsingou (FPUT) lattice \cite{fermi1955studie}, a chain of modes satisfying the selection rule can be established and shape the thermalization route \cite{Flach2005PRL,Iva2009PRL,Bivins1973}. Another distinct mechanism is through Mathieu instability \cite{Bivins1973, induction_1d, induction_mathieu}. Moreover, when the nonlinearity is strong enough, the resonant peaks of the modes are broadened, leading to the Chirikov resonance mechanism for thermalization \cite{energy_equipartition1,Lov2018PRL}. {\color{black} Most of these recent theoretical works focused on one dimensional (1D) nonlinear lattices, only a few examined thermalization in two dimensional (2D) cases \cite{twodimensional_FPU1,twodimensional_FPU2,Zhao2D2020}.} For graphene resonators, due to the complex interatomic interaction, the mode couplings are much more complicated than that in the 1D FPUT lattice \cite{Luo2021NL,Math2016NN,Eich2012PRL,math2013nonlinear, Westra2010PRL}.

A key quantity characterizing the thermalization process is the equipartition time $\tau$, which is the time needed for the energy initially localized on a mode, i.e. the initially excited mode (IEM), to be equally distributed among all the modes \cite{WT_pnas,Zhao2020PRL,Lov2018PRL}.
The equipartition time is in general reversely related with the intermodal energy transfer rate that could be induced by the internal resonance \cite{Luo2021NL,Math2016NN,Eich2012PRL} or nonlinear mode coupling \cite{math2013nonlinear,Westra2010PRL}. A common observation is that the equipartition time gets shorter as the energy of the system becomes larger. Indeed, this has been widely corroborated in previous investigations of {\color{black} 1D \cite{WT_pnas,Zhao2020PRL,Lov2018PRL,Dhar2007PRL,Metastability_scenario, FPU_rev} or 2D \cite{FPUprl,wang2018symmetry,wang2018pre,twodimensional_FPU1,twodimensional_FPU2,Zhao2D2020}} nonlinear lattices. In particular, the specific scaling form of $\tau$ versus the initial excitation energy is either stretched exponential
\cite{FpuNekhoroshev1,FpuNekhoroshev2, FpuNekhoroshev_support,Toda_paper2} or power law \cite{Lov2018PRL,Fu2019PRE,Fu2019NJP,time_scale_power,benettin2013fermi}. For both cases, larger initial energy would infer shorter equipartition time.

In this Letter, we investigate the thermalization of a circular graphene nanoresonator with molecular dynamics (MD) simulations and unveil a {\color{black} thermalization frustration} phenomenon: for many normal modes as the IEM, there exists an energy range, that as the energy increases, instead of being shorter, the equipartition time increases abruptly and can be an order longer. Thus larger energy may even slow down significantly the equipartition process.
{\color{black} This phenomenon can be associated with a dynamical instability, i.e., the Mathieu instability} between different classes of modes partitioned by the hierarchical symmetry, and may lead to dynamical insights in the internal energy dissipation mechanism of graphene nanoresonators. Since nanoresonators are formed by perfect atomic lattices with little dislocations, the MD simulation results with realistic potential field are expected to conform with the experiments \cite{SmirnovSM2014}.

\begin{figure}[t]
\centering
\includegraphics[width=\linewidth]{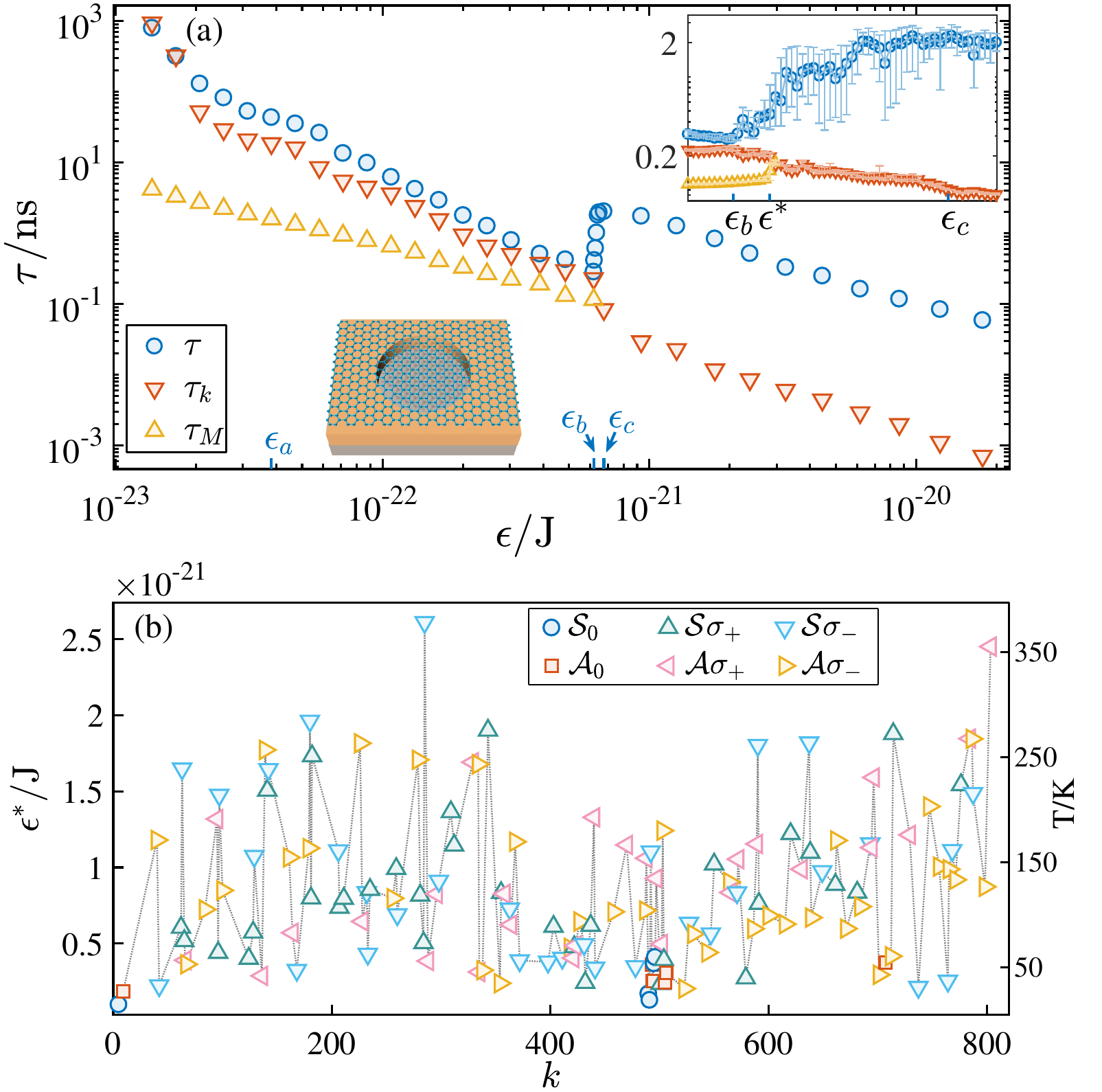}
\caption{{\bf (a)} Dependence of the {\color{black} ensemble average of the} equipartition time $\tau$ and two characteristic timescales $\tau_k$ and $\tau_M$ on specific energy $\epsilon$ when the IEM $k= 437$.
The upper inset zooms in the region $[\epsilon_b, \epsilon_c]$ {\color{black} with error bars} and shows the thermalization frustration phenomenon. {\color{black} The statistics are obtained from 100
uniformly distributed phases in $[0,2\pi]$.
The lower inset shows the schematic plot of the resonator.} {\bf (b)} The dependence of $\epsilon^*$ on the mode index $k$ for those who experience thermalization frustration {\color{black} with $\phi = 0$}. The corresponding temperatures are shown in the right $y$-axis. {\color{black} The computation of these data, without ensemble statistics, takes about 7 months with on average 100 cores (Intel(R) Xeon(R) Gold 5218 CPU @2.30GHz).}}
\label{fig:equ_time}
\end{figure}

{\it Model.}---
We consider a single-layer circular graphene resonator \cite{Bunch2007,lee2008,Guttinger2017NN,NC_tian2018} {\color{black} as sketched in the lower inset of Fig. \ref{fig:equ_time}(a).} The boundary atoms are fixed, only the inner sites can move in the force field. In our simulation the diameter of the resonator is approximately 7.95 nm, with $N=1884$ movable sites.
For mechanical resonators, the in-plane motions {\color{black} inside the graphene plane} are greatly suppressed, and the out-of-plane {\color{black} oscillations perpendicular to the plane} are their most dominant dynamics \cite{FPUprl,science_seol2010}. This leads to a simplified potential of the valence force field for the $sp^2$ bond in graphene \cite{VFF_model,ForceModel1,ForceModel2,VFFmodel2,wang2018pre}:
{\color{black}
\begin{equation}
\begin{split}
&U=\sum_{i=1}^{N}\bigg[\gamma(\sum_{j\in N_i}z_j-3z_i)^2+\frac 12\sum_{j\in N_i}\frac \alpha{4a_0^2}(z_j-z_i)^4+\\&\sum_{\substack{j,l\in N_i\\j<l}}\frac \beta{a_0^2}[(z_j-z_i)(z_l-z_i)]^2\bigg]\equiv U^{(2)}+U^{(4)}_1+U^{(4)}_2,
\end{split}
\label{equ:poten_sim}
\end{equation}
where $i$ is the site index, $N_i$ is the set of $i$'s nearest neighbors,} $z_i$ is the $z$-displacement from the equilibrium position, and can be written in a vector form ${\bf z}(t)=[z_1(t),\cdots,z_N(t)]^T$, $a_0=1.421$ $\mathring{\mbox{A}}$ is the equilibrium bond length [Supplemental Material (SM) Sec. I].
The parameters $\alpha$, $\beta$, and $\gamma$ are 155.9, 25.5, and 7.4 $\mbox{J}/\mbox{m}^{2}$ to account for realistic atomic interactions \cite{VFF_model,VFFmodel2}. The normal modes and their corresponding angular frequencies {\color{black} in increasing order,} $\{\varphi_i, \omega_i, i=1,\cdots,N\}$, can be obtained by diagonalizing the stiffness matrix derived from the second order potential $U^{(2)}$ \cite{V2U2,wang2020PRB}.

The thermalization process is investigated with MD simulations according to potential (\ref{equ:poten_sim}). Verlet algorithm is employed with a time step of 0.5 fs to integrate the configuration profile, {\color{black} which can be expanded to the orthogonal normal modes as ${\bf z}(t) = \sum_i c_i(t) \varphi_i$.} The harmonic energy of each mode is given by $E_i(t)=\frac 12m_{\rm c}(\dot {c}_i^2+\omega_i^2c_i^2)$ \cite{FPU_rev,multistage}, $m_{\rm c}$ being the mass of carbon atom.
{\color{black} Initially, the graphene sheet is perturbed according to mode $k$ (the IEM) with initial harmonic energy $E$ and phase $\phi$ \cite{WT_pnas,Lov2018PRL}: $c_k(0)=(\sqrt{2 E /m_{\rm c}}/\omega_k)
\cos(\phi)$, $\dot{c}_k(0)= c_{\phi} \sqrt{2 E /m_{\rm c}}
\sin(\phi)$, while $c_i(0)=\dot{c}_i(0)=0$ for $i\neq k$.
The specific energy $\epsilon$ per mode including the nonlinear potential is obtained with $\phi=0$, and is unchanged ensured by the rescaling factor $c_{\phi}$ for different $\phi$.}
The process toward thermalization is characterized by $\sigma(t)=\langle[\log_{10}E_i(t)-\langle\log_{10}E_i(t)\rangle]^2\rangle^{1/2}$, where the average is over all the normal modes. The equipartition time $\tau$ is defined as $\sigma(\tau)=0.9$ {\color{black} given that $\tau$ is larger than the time when $\sigma(t)$ maximizes} (SM Sec. II) \cite{wang2018symmetry}.

\begin{figure*}[t]
\centering
\includegraphics[width=1.01\linewidth]{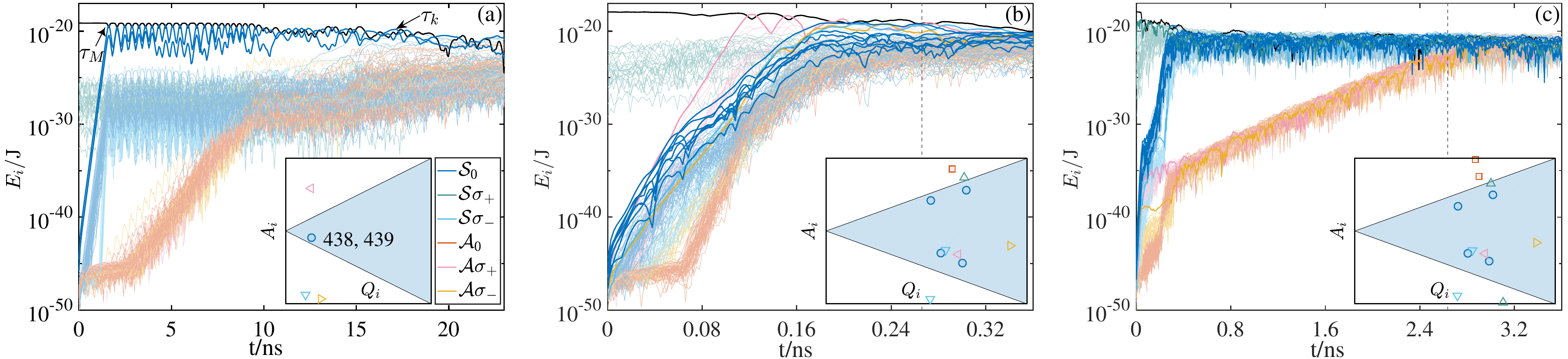}
\caption{The time evolution of the harmonic energies for all the modes. The IEM $k=437$ {\color{black} with $\phi = 0$} (the thick black line). In order to highlight the modes with Mathieu instability, only the unstable modes are drawn with thick lines, while the other modes are plotted with thin lines. {\bf (a)} $\epsilon_a=3.836\times 10^{-23}$ J, and the equipartition time $\tau = 42$ ns.
{\bf (b)} $\epsilon_b=6.198\times 10^{-22}$ J, and $\tau = 0.27$ ns.
{\bf (c)} $\epsilon_c=6.567\times 10^{-22}$ J, and $\tau = 2.63$ ns.
Note that the time window for (c) is exactly an order longer than that for (b). The insets show the resonance region (the blue shaded region) of the Mathieu equation, and the unstable modes that fall into this region at the corresponding energies.}
\label{fig:energy_flow}
\end{figure*}

{\it Phenomenon and mechanism.}---
As an exemplary case, Fig. \ref{fig:equ_time}(a) plots the equipartition time $\tau$ (the blue circles) versus the specific energy $\epsilon$ when the IEM is 437 (see SM Sec. III for more cases).
{\color{black} Three representative energy values, $\epsilon_a$, $\epsilon_b$, and $\epsilon_c$, are marked and will be investigated further in Fig. \ref{fig:energy_flow}.}
When $\epsilon$ passes $\epsilon_b$, $\tau$ increases abruptly along with big fluctuations and rises to an order longer around $\epsilon_c$. Therefore, as the initially injected energy increases, the system needs longer time to reach thermalization. As the general trend of $\tau$ versus $\epsilon$ in a much larger energy scale is decreasing, this {\color{black} sudden increase} indicates a frustration of the thermalization process. {\color{black} Despite wild fluctuations, the phenomenon is robust against ensemble statistics [upper inset of Fig. \ref{fig:equ_time}(a), see also SM Sec. IV].}

Additionally, Fig. \ref{fig:equ_time}(a) shows two key timescales before equipartition is arrived.
One is $\tau_k$ when the IEM $k$'s energy $E_k$ drops significantly that this mode loses its dominant role in the thermalization dynamics, i.e., $\frac{1}{\tau_k} \int_0^{\tau_k}E_k(t)dt = 0.5 E_k(0)$.
The other is $\tau_M$ when the energy $E_M$ of the mode with Mathieu instability becomes comparable with $E_k$ without interruption, i.e., $E_{M}(\tau_M) = 0.5 E_k(0)$.
{\color{black} Since mode $k$ supplies the driving source to all the other modes, $\tau_M$ is meaningful only when $\tau_M < \tau_k$, that the IEM $k$ is still dominating at $\tau_M$.}
For small energies, $\tau_k$ is larger than $\tau_M$.
As $\epsilon$ increases, both $\tau_k$ and $\tau_M$ decrease, but $\tau_k$ drops faster. The two intersect at a certain point, e.g., $\epsilon^*\sim 6.26\times 10^{-22}$ J, marking the characteristic energy scale for the thermalization frustration phenomenon.
{\color{black} The maximum eigenfrequency in the acoustic branch is $f_{942}=\omega_{942}/2\pi=13.758$ THz.
Since $f_{437}= 6.444$ THz, this mode locates around the middle between the zone center and boundary in the wavevector space.}

{\color{black} This phenomenon is abundant.
In 260 modes that are randomly selected as the IEM from the 942 acoustic modes of this system, more than half (148 modes) experience thermalization frustration.
In addition, this phenomenon is persistent for systems with different sizes, and has also been identified in a more realistic REBO potential based simulation considering all the $x,y,z$ motions (SM Sec. V).}
Figure \ref{fig:equ_time}(b) summarizes the dependence of $\epsilon^*$ on the mode index $k$ of these 148 modes.
Regarding $\epsilon^*$ as the energy of the thermal motion, the corresponding temperatures are shown in the right $y$-axis, which are approximately in the range of 20 K to 370 K. Thus the phenomenon can be expected to be observable for those modes with large $\epsilon^*$, as their energy scale is substantially higher than the thermal noise in typical nanoresonator experiments \cite{Eich2012PRL}.

\begin{figure}[t]
\centering
\includegraphics[width=\linewidth]{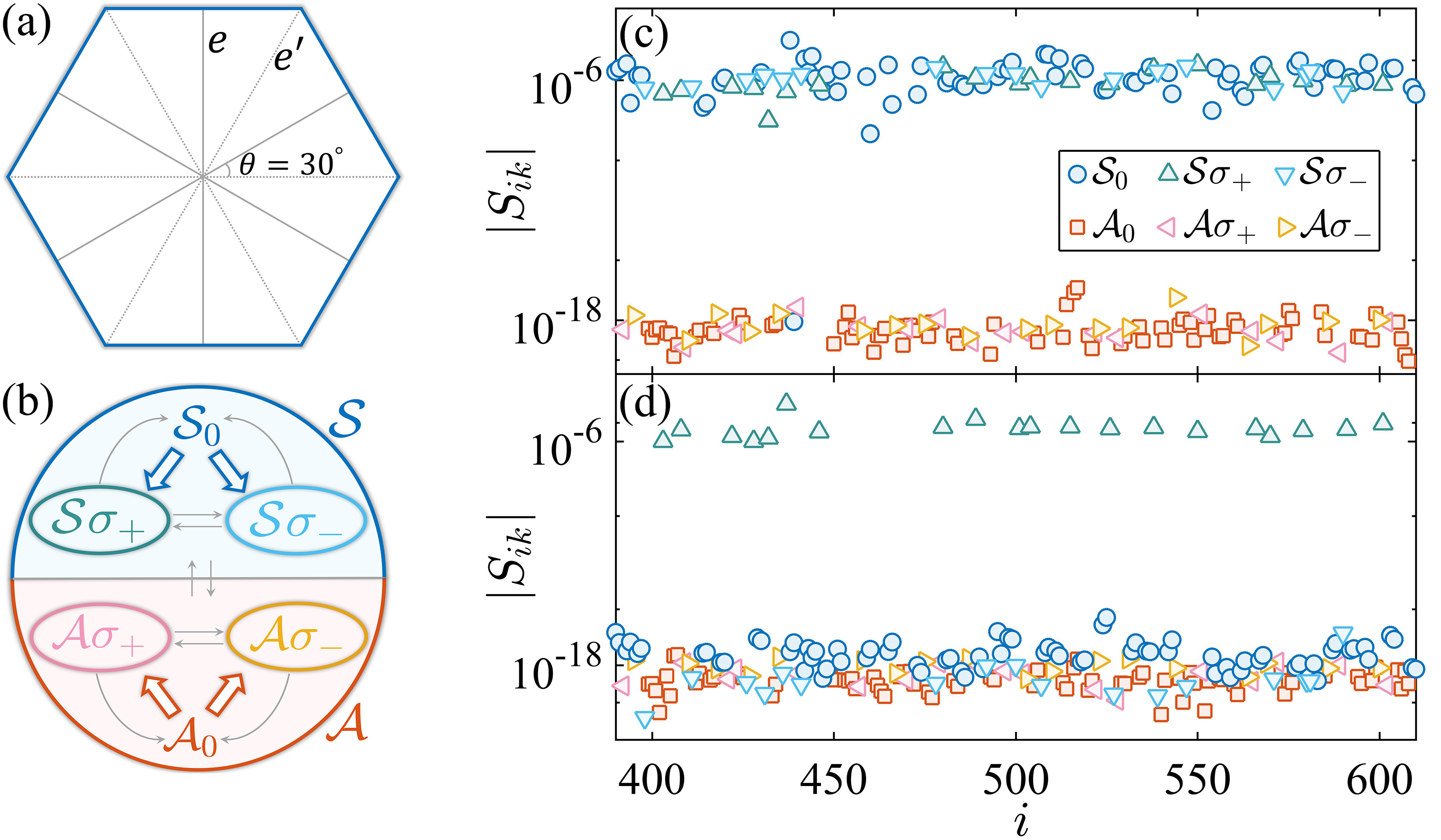}
\caption{{\bf (a)} The mirror and rotational symmetries of the honeycomb lattice.
{\bf (b)} The diagram showing the hierarchical structure of the symmetry classes. The big arrows indicate strong couplings that are in the same order as the intra-class couplings, and the thin gray arrows represent negligible couplings. There are 628, 166, and 148 modes in $\mathcal{S}_0$ ($\mathcal{A}_0$), $\mathcal{S\sigma}_+$ ($\mathcal{A\sigma}_+$), and $\mathcal{S\sigma}_-$ ($\mathcal{A\sigma}_-$), respectively.
{\bf (c, d)} The coupling strength $|S_{ik}|$ for $k= 438 \in \mathcal{S}_0$ (c) and $k= 437 \in \mathcal{S\sigma}_+$ (d).
}
\label{fig:model}
\end{figure}

The mechanism lies in the spatiotemporal dynamics of the intermodal energy transfer, which depends on the symmetry based partition of the normal modes.
The system has mirror symmetries along two sets of axes, one connects the origin to the center of the bond (denoted by $e$), the other connects the origin to the atom ($e'$), as indicated in Fig. \ref{fig:model}(a), together with 6-fold rotational symmetry ($c_6$), which automatically ensures 2-fold ($c_2$) and 3-fold ($c_3$) rotational symmetries.
Since the normal modes are real, one has $c_m \varphi_i=\pm\varphi_i$, $m=2,3,6$ \cite{Refc3}. All the modes have $c_2$ symmetry, half with $c_2\varphi_j=\varphi_j$, denoted by $\mathcal{S}$, and the other half with $c_2\varphi_j=-\varphi_j$, denoted by $\mathcal{A}$.
For each half, there are modes that do not follow $c_6$, which are denoted as $\mathcal{S}_0$ and $\mathcal{A}_0$. For those who follow $c_6$ and in $\mathcal{S}$, one must have $c_6\varphi_i=\varphi_i$.
Under mirror reflection $\sigma_e$ ($\sigma_{e'}$) with respect to axes $e$ ($e'$), they are either symmetric or anti-symmetric under both $\sigma_e$ and $\sigma_{e'}$, which are denoted as $\mathcal{S\sigma}_+$ and $\mathcal{S\sigma}_-$, respectively. Thus $\mathcal{S}=\mathcal{S}_0+\mathcal{S\sigma}_++\mathcal{S\sigma}_-$.
Similarly, $\mathcal{A} = \mathcal{A}_0+\mathcal{A\sigma}_++\mathcal{A\sigma}_-$ \cite{RefA}.
Therefore the normal modes can be grouped into six symmetry classes (SM Sec. I). The intermodal coupling, as determined by the nonlinear terms in Eq. (\ref{equ:poten_sim}), can be characterized by the coupling strength
$S_{ij}=\varphi_i^{\dagger}{\bf V}^{\text{(4)}}\varphi_j$, where $V_{pq}^{\text{(4)}}\equiv-\frac 1{3}\frac{\partial^2 (U^{(4)}_1+U^{(4)}_2)}{\partial z_p\partial z_q}$ \cite{wang2018pre}. In general, within a class, all the modes are strongly coupled. Between different classes, $S_{ij}$ is generally ten orders smaller, which is effectively zero, preventing energy flow \cite{wang2018symmetry}. {\color{black} There can be other possible mechanisms classifying the modes. For example, in a 1D FPUT system, the modes have been grouped by either short or long wavelengths, with negligible interactions between the two classes \cite{Rondonietal2020}.}

However, particularly for the circular graphene resonator, the six symmetry classes are nonequivalent, leading to an interesting asymmetric coupling and forming a symmetry hierarchy, as shown in Fig. \ref{fig:model}(b). To be specific, if the IEM $k$ belongs to $\mathcal{S}_0$, then it ``sees'' all the modes in $\mathcal{S}$, including those in $\mathcal{S\sigma}_+$ and $\mathcal{S\sigma}_-$, as in the ``same'' class, that the coupling $S_{ik}$s are all in the same order of a large magnitude [Fig. \ref{fig:model}(c)]. However, if $k$ belongs to $\mathcal{S\sigma}_+$ (or $\mathcal{S\sigma}_-$), then only for $i$ in the same class, $S_{ik}$ can take a large value [Fig. \ref{fig:model}(d)]. The same asymmetry occurs between $\mathcal{A}_0$ and $\mathcal{A\sigma}_+$ ($\mathcal{A\sigma}_-$) \cite{RefSikS}.
Furthermore, $S_{ik}$s between the $\mathcal{S}$ and $\mathcal{A}$ classes are zero as they are even and odd under $c_2$, respectively.
$S_{ik}$s between $\mathcal{S\sigma}_+$ and $\mathcal{S\sigma}_-$, or $\mathcal{A\sigma}_+$ and $\mathcal{A\sigma}_-$, are also zero because their parity is opposite.

This leads to distinct modal dynamics. Figure \ref{fig:energy_flow}(a) shows the time evolution of the harmonic energies of all the modes when the IEM $k=437 \in \mathcal{S\sigma}_+$ with a representative specific energy $\epsilon_a$ as marked in Fig. \ref{fig:equ_time}(a). For those modes in $\mathcal{S\sigma}_+$, $S_{ik}$ is large [Fig. \ref{fig:model}(d)], then mode $i$ can be approximated by a driven harmonic oscillator \cite{wang2018pre}, i.e., $m_{\rm c}\ddot{c}_i\approx - m_{\rm c}\omega_i^2c_i+S_{ik}c_k$, with the last driving term from the IEM. As a result, these modes will gain energy almost instantaneously (just in a few steps), but typically they are still a few orders smaller than that of the IEM, as shown in Fig. \ref{fig:energy_flow}. Since the solution of the driven oscillator is stable, their energy levels are stabilized, as indicated by the almost flat dark green curves.

However, if mode $i \notin \mathcal{S\sigma}_+$, $S_{ik}$ can be ten orders smaller [Fig. \ref{fig:model}(d)], thus the direct driving can be neglected. However, there are higher order terms that enter into the modulation of the frequency, resulting in parametric resonance following the Mathieu equation $\ddot{c}_i+\big[A_i-2Q_i\mbox{cos}(2t)\big]c_i=0$ (SM Sec. VI) \cite{wang2018pre}. Note that $(Q_i,A_i)$ depends on the selected IEM $k$ and also the initially injected energy.
The inset of Fig. \ref{fig:energy_flow}(a) shows the resonance region of the Mathieu equation, and the parameter pairs $(Q_{i}, A_{i})$ for modes 438 and 439 in class $\mathcal{S}_0$ fall in this region. Thus these two modes are unstable and gain energy in an exponential way, as demonstrated by the thick blue lines. Due to strong couplings of the modes in $\mathcal{S}_0$ to all the modes in $\mathcal{S}$, the other modes in $\mathcal{S}_0$, together with the modes in $\mathcal{S\sigma}_-$, are all lifted up by these two modes. For class $\mathcal{A}$ (the three warm color lines), as they do not have an unstable mode, their energies remain negligible small until around $t=3.0$ ns when all the $\mathcal{S}$ modes are lifted up. These modes gain energy in a much slower way than that induced by Mathieu instability, resulting in a long equipartition time for the whole system. This process is typical and occurs for many randomly chosen IEMs.

To unveil the dynamical origin of the thermalization frustration, we choose the two values, $\epsilon_b$, smaller but close to $\epsilon^*$, and $\epsilon_c$, as marked in Fig. \ref{fig:equ_time}(a), and plot the corresponding time evolution of the harmonic energies of the modes in Fig. \ref{fig:energy_flow}(b) and \ref{fig:energy_flow}(c), respectively.
Although $\epsilon_c$ is larger, the corresponding equipartition time $\tau$ is approximately an order longer than that for $\epsilon_b$.
This is due to the sudden change in the energy flow pathways around $\epsilon^*$. For $\epsilon_b$, the two characteristic timescales are approximately equal, i.e., $\tau_k\sim\tau_M$. The unstable modes have just enough time to get energy exponentially fast until their energies are comparable with $E_k$, as indicated by the thick lines in Fig. \ref{fig:energy_flow}(b), which results in a short equipartition time. While for $\epsilon_c$, in the beginning, the IEM $k$ is dominant and drives the modes with Mathieu instability that they gain energy in an exponential way. However, around $t \sim\tau_k =0.084$ ns, due to the drop of its energy, the IEM loses the dominant role. Without the driving source, the Mathieu instability mechanism terminates, and the fast increase of the energy for these modes stops. After that, although these modes may still gain energy in an exponential way, as shown in Fig. \ref{fig:energy_flow}(c), the slope is much smaller, resulting in an overall much longer equipartition time. {\color{black} Here the thermalization within each class is fast, thus before $\tau$, different classes may have different effective temperatures.}

An additional feature of the equipartition time in the energy range $[\epsilon_b,\epsilon_c]$ is that, as $\tau$ increases, it experiences huge fluctuations [upper inset of Fig. \ref{fig:equ_time}(a)]. For these cases, after $\tau_k$, the original Mathieu instability terminates.
However, there may appear a second process of Mathieu instability, when the IEM occasionally recovers (partially) its dominant role. Alternatively, a different mode may become dominant, and a different set of modes may fall in the resonance region caused by the new dominant mode (SM Sec. VII). When this happens, the equipartition time can be significantly reduced. The condition for this to happen is very sensitive, that a tiny variation of the initially injected energy {\color{black} or the initial phase} may lead to or destroy such processes, resulting in huge fluctuations in the final equipartition time.

{\it Conclusion and discussion.}---
A thermalization frustration phenomenon has been unveiled with extensive MD simulations in graphene nanoresonators. Despite the overall trend that the equipartition time decreases with increasing energy, there may exist an energy range that the equipartition time can increase abruptly along with huge fluctuations. And it can be an order longer. The underlying mechanism is the creation, termination, and possible recreation (with sensitive energy dependence) of the Mathieu instability, which opens, closes, and reopens the energy flow channels between different classes partitioned by the hierarchical symmetry structure in the normal modes.
These results cast profound new understandings to thermalization dynamics, i.e., the thermalization process at nanoscale may rely heavily on the dynamics dominated by only a few pivotal degrees of freedom.

Thanks to the state-of-the-art technical advances in manipulating graphene nanoresonators \cite{yg_sensor,Guttinger2017NN} and direct measurement of phonon lifetime \cite{SongWD2008,WangSG2010}, the phenomenon might be observed directly in nanoelectromechanical resonator experiments. The mechanism is expected to hold for larger systems, which can be experimentally more tractable. Recent developments of FPUT physics in photonic lattices \cite{Christodoulides2003, LahiniAP2008,kon2015phot,Parity2019}, trapped-ion crystals \cite{RMP2019many,Detail2018,Time2016clos,LemmerCSS2015,DingMHM2017} and optical fibers \cite{muss2018fibre,Obser2018,Exp2007} provide additional experimental platforms that could be exploited to investigate the phenomenon. From the application point of view, energy dissipation of the dominant mode ties in closely with the relaxation dynamics and the quality factor \cite{Wilson2011PRL}, thus our results may provide controlling strategies for these systems.

We thank Prof. Hong-Ya Xu and Dr. Zhigang Zhu for helpful discussions.
This work was supported by NSFC under Grant Nos. 11905087, 12175090, 11775101, and 12047501, and by NSF of Gansu Province under Grant No. 20JR5RA233.

\bibliography{ref}

\begin{thebibliography}{73}
\bibitem [{V2U()}]{V2U2}%
  \BibitemOpen
  \bibinfo {note} {{\color{black} The stiffness matrix ${\bf V}^{\text{(2)}}$
  can be derived from the second order potential $U^{(2)}$ as
  $V_{ij}^{(2)}=-\frac{\partial^2 U^{(2)}} {\partial z_i\partial z_j}$,
  yielding $V_{ij}^{(2)}=-24\gamma$ if $j=i$, $V_{ij}^{(2)}=12\gamma$ if $j$ is
  $i$'s nearest neighbor, and $V_{ij}^{(2)}=-2\gamma$ if $j$ is $i$'s next
  nearest neighbor, and $V_{ij}^{(2)}=0$ otherwise. Assume $\lambda_i$ is the
  eigenvalue of ${\bf V}^{\text{(2)}}$, then the corresponding angular
  frequency is $\omega_i=\sqrt{-\lambda_i/m_{\rm c}}$.}}\BibitemShut {Stop}%
\bibitem [{Ref({\natexlab{a}})}]{Refc3}%
  \BibitemOpen
  \bibinfo {note} {It can be demonstrated that for $c_3$, the case
  $c_3\varphi_i=-\varphi_i$ is prohibited. Furthermore, all $\varphi_i$s that
  satisfy $c_3 \varphi_i = \varphi_i$ must also have the 6-fold rotational
  symmetry: $c_6\varphi_i=\pm\varphi_i$. While for $c_2$, there are some modes
  $\varphi_j$ that only satisfy $c_2$, i.e., $c_2\varphi_j=\pm\varphi_j$, but
  not $c_6$.}\BibitemShut {Stop}%
\bibitem [{Ref({\natexlab{b}})}]{RefA}%
  \BibitemOpen
  \bibinfo {note} {For those who follow $c_6$ but in $\mathcal{A}$, one must
  have $c_6\varphi_i=-\varphi_i$, and they are symmetric (anti-symmetric) with
  respect to $\sigma_e$ ($\sigma_{e'}$), denoted as $\mathcal{A\sigma}_+$, or
  anti-symmetric (symmetric) with respect to $\sigma_e$ ($\sigma_{e'}$),
  denoted as $\mathcal{A\sigma}_-$.}\BibitemShut {Stop}%
\end{thebibliography}

\end{document}